\documentclass[english]{sbrt}
\usepackage{float}
\usepackage[english]{babel}
\usepackage[utf8]{inputenc}
\usepackage{lipsum}
\usepackage{graphicx}
\usepackage{multirow}
\usepackage{pifont}
\usepackage{threeparttable}
\usepackage{cite}
\usepackage{url}
\usepackage{fancyhdr}

\begin{document}

\title{Towards Network Data Analytics in 5G Systems and Beyond}

\author{Marcos Lima Romero and Ricardo Suyama
\thanks{Marcos Lima Romero, e-mail: marcos.romero@ufabc.edu.br; Ricardo Suyama, e-mail: ricardo.suyama@ufabc.edu.br; CECS, UFABC, Santo André-SP. This work was partially supported by Coordenação de Aperfeiçoamento de Pessoal de Nível Superior - Brasil (CAPES) -- Financial Code 001, by Fundação de Amparo à Pesquisa do Estado de São Paulo (FAPESP) - grant \#2020/09838-0, and the Conselho Nacional de Desenvolvimento Científico e Tecnológico (CNPq) - grant \#311380/2021-2.}%
}

\maketitle

\begin{abstract}
Data has become a critical asset in the digital economy, yet it remains underutilized by Mobile Network Operators (MNOs), unlike Over-the-Top (OTT) players that lead global market valuations. To move beyond the commoditization of connectivity and deliver greater value to customers, data analytics emerges as a strategic enabler. Using data efficiently is essential for unlocking new service opportunities, optimizing operational efficiency, and mitigating operational and business risks. Since Release 15, the 3rd Generation Partnership Project (3GPP) has introduced the Network Data Analytics Function (NWDAF) to provide powerful insights and predictions using data collected across mobile networks, supporting both user-centric and network-oriented use cases. However, academic research has largely focused on a limited set of methods and use cases, driven by the availability of datasets, restricting broader exploration. This study analyzes trends and gaps in more than 70 articles and proposes two novel use cases to promote the adoption of NWDAF and explore its potential for monetization. 

\end{abstract}
\begin{keywords}
NWDAF, Data Analytics, Machine Learning, 5G, 3GPP, Survey
\end{keywords}

\section{Introduction}

With the evolution of mobile networks, particularly 5G, Mobile Network Operators (MNOs) are handling vast amounts of data daily, which could serve as the foundation for advanced Artificial Intelligence (AI) and Machine Learning (ML) models. However, operators are often overwhelmed by Over-the-Top (OTT) applications and struggle to take advantage of these data. Many MNOs continue to treat their services as commodities, focusing solely on connectivity and leaving the added value to OTT providers \cite{kim2025relieve}. This approach limits their competitiveness in the face of digital transformation and highlights the need for a significant shift in data management and network automation. Using data analytics, MNOs could introduce new value-added services to end users, unlock additional business opportunities, and mitigate operational risks.

Since Release 15 in 2017 \cite{3gpp-23501-rel15}, the 3rd Generation Partnership Project (3GPP) has been exploring new ways to enhance data analysis in the 5G system. The Network Data Analytics Function (NWDAF) was proposed as a solution to centralize data collection and ML models within the Core Network, with the goal of providing analytics and predictions to other network functions (NFs), application functions (AFs), and operation and maintenance (OAM) functions. NWDAF has evolved over successive releases, starting with a single use case focused on network slicing support, and now includes more than 23 use cases in the latest Release 19 \cite{3gpp-23288-rel19}.

Given this challenging scenario, the main contributions of this paper are as follows:
\begin{itemize} 
    \item Present the latest advancements in 3GPP standards related to network automation within the Core Network. 
    \item Summarize the current academic state-of-the-art in the development and implementation of the NWDAF. 
    \item Analyze how researchers are addressing standardization challenges, exploring current use cases, and examining the algorithms being applied. 
    \item Identify gaps in the existing literature and propose novel use cases to promote the adoption of NWDAF and explore its monetization potential. 
\end{itemize}

The following sections are organized as follows: Section \ref{sec:3GPP} presents the evolution of NWDAF through 3GPP releases. Related works are discussed in Section \ref{sec:related}. The methodology used to search for relevant papers is outlined in Section \ref{sec:methodology}. Sections \ref{sec:results} and \ref{sec:proposed} discuss the results of the literature review and identify its gaps and propose new use cases. Finally, Section \ref{sec:conclusion} concludes the paper and provides future research directions.

\section{NWDAF Role in 3GPP \label{sec:3GPP}}

NWDAF started as a concept in the first 3GPP 5G release, Release 15, and was first introduced in the 5G system architecture specification in TS 23.501 \cite{3gpp-23501-rel15}. The initial use case for NWDAF involves interaction with the Policy Control Function (PCF) to provide information about the network slice load level. This service was later detailed in 2018 in TS 29.520, also in Release 15 \cite{3gpp-29520-rel15}.

The architecture for NWDAF was first discussed in 2019 within the 
{Enablers for Network Automation of 5G (eNA) work of the System Architecture working group (SA2), leading to TS 23.288 \cite{3gpp-23288-rel16} in Release 16. The signalling flow was introduced later, in 2022, with Release 17 in TS 29.552 \cite{3gpp-29552-rel17}. In the Service-Based Architecture (SBA), NWDAF was defined to operate using subscribe-notify or request-response modes. The use cases also evolved over the years, as shown in Fig. \ref{fig:3gpp}. In Release 15, there was one use case; in Release 16, 10 use cases \cite{3gpp-23288-rel16}; in Release 17, 15 use cases \cite{3gpp-23288-rel17}; in Release 18, 21 use cases \cite{3gpp-23288-rel18}; and, as of now, Release 19, 23 use cases \cite{3gpp-23288-rel19}. For ongoing Release 20, a new use case is suggested. Table \ref{tab:use_cases} shows the use cases and a segregation into four main categories: Slice, Network, Service, and User.

\begin{figure*}
    \centering
    \includegraphics[width=.79\linewidth]{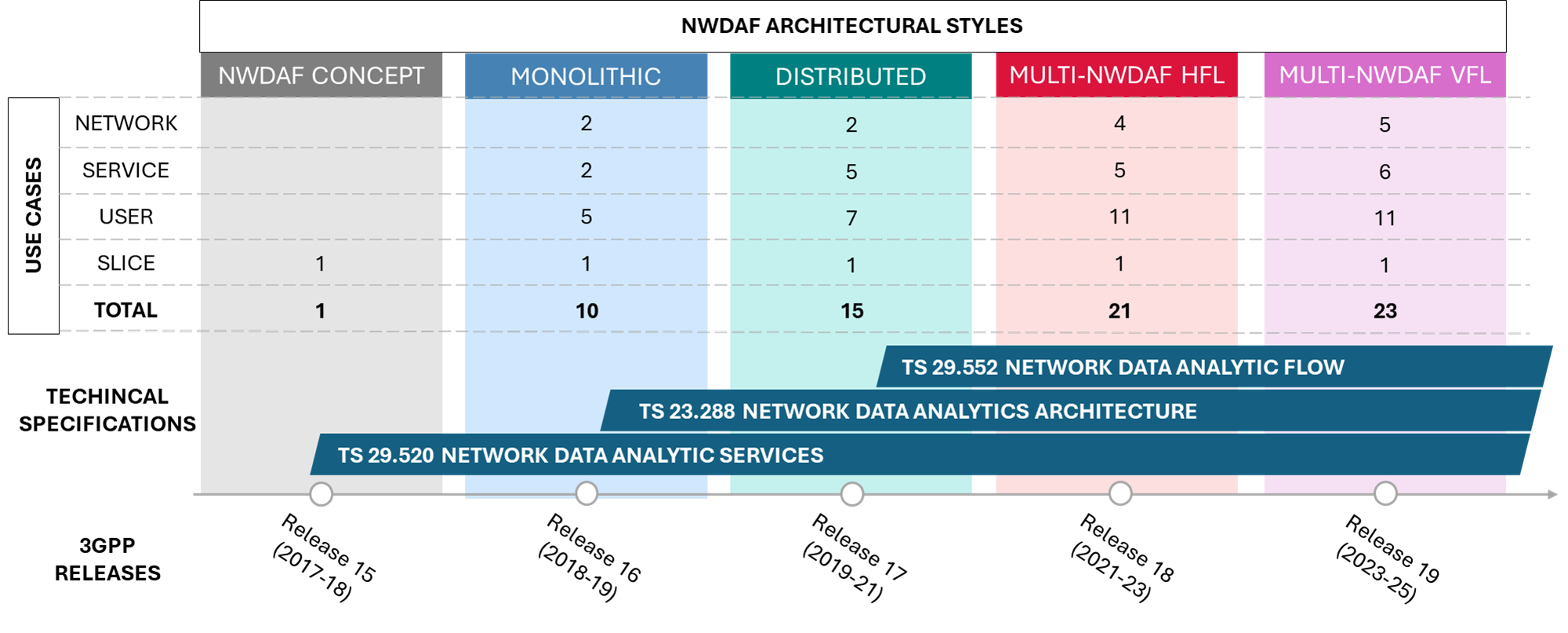}
    \caption{Evolution of NWDAF over 3GPP Releases.}
    \label{fig:3gpp}
\end{figure*}

\begin{table}[hbt]
\centering
\caption{Use cases by category and 3GPP Release}
\label{tab:use_cases}
\resizebox{\columnwidth}{!}{%
\begin{tabular}{|c|c|c|}
\hline
\textbf{Release} & \textbf{Category} & \textbf{Use Case} \\ \hline
15 & Slice & Slice Load level information \\ \hline
\multirow{9}{*}{16} & \multirow{2}{*}{Network} & NF Load Information \\ \cline{3-3}
 &  & Network Performance information \\ \cline{2-3}
 & \multirow{2}{*}{Service} & Observed Service Experience \\ \cline{3-3}
 &  & QoS Sustainability \\ \cline{2-3}
 & \multirow{5}{*}{User} & UE mobility information \\ \cline{3-3}
 &  & UE communication information \\ \cline{3-3}
 &  & Expected UE behavioural parameters \\ \cline{3-3}
 &  & UE Abnormal behaviour information \\ \cline{3-3}
 &  & User Data Congestion information \\ \hline
\multirow{5}{*}{17} & \multirow{3}{*}{Service} & \begin{tabular}[c]{@{}c@{}}Session Management Congestion Control\\ Experience\end{tabular} \\ \cline{3-3}
 &  & Redundant Transmission Experience \\ \cline{3-3}
 &  & DN Performance \\ \cline{2-3}
 & \multirow{2}{*}{User} & WLAN Performance \\ \cline{3-3}
 &  & Dispersion \\ \hline
\multirow{6}{*}{18} & \multirow{2}{*}{Network} & PFD Determination \\ \cline{3-3}
 &  & PDU Session Traffic \\ \cline{2-3}
 & \multirow{4}{*}{User} & End-to-end data volume transfer time \\ \cline{3-3}
 &  & Movement Behaviour \\ \cline{3-3}
 &  & Location Accuracy \\ \cline{3-3}
 &  & Relative Proximity \\ \hline
\multirow{2}{*}{19} & Network & Signalling Storm \\ \cline{2-3}
 & Service & QoS and Policy Assistance \\ \hline
20 & Network & Abnormal UPF traffic pattern \\ \hline
\end{tabular}%
}
\end{table}

The Slice category includes the earliest use case, in which the PCF requests slice load level information. The Network category comprises use cases related to statistical analysis and prediction 
of NF loads, as well as performance metrics associated with Access Network resource utilization. In Release 18, two additional use cases were introduced under this category: the determination of Packet Flow Descriptions (PFDs) and the analysis of Packet Data Unit (PDU) session traffic associated with User Equipment (UE). Furthermore, Release 19 expands this category by introducing use cases for predicting signalling storms, with the goal of preventing abnormalities.

The Service category encompasses use cases focused on Quality of Service (QoS) and Quality of Experience (QoE). These include observed service experience as delivered by a specific slice or application, QoS sustainability, characterized by changes in QoS levels across time and location, along with associated thresholds, and session management congestion control for specific Data Network Names (DNNs) or slices. Additional scenarios include redundant transmission experiences for Ultra-Reliable Low Latency Communication (URLLC) services, user plane performance evaluation for Edge Computing applications, and QoS and policy assistance mechanisms to support the selection of optimal QoS parameter sets aligned with expected QoE requirements.

The User category includes use cases centered on UE-related behaviours and performance. In Release 16, use cases were defined to address UE mobility and communication patterns, as well as detection of expected and abnormal behaviour and user data congestion. Release 17 introduced use cases concerning WLAN performance metrics for UEs, as well as the dispersion of UEs with respect to data volume, mobility patterns, and session management transactions. Release 18 extended this scope further by incorporating use cases related to end-to-end data volume transfer time statistics or predictions between UEs and applications, movement behaviour, location accuracy, and relative proximity between UEs.

Currently, 3GPP is starting the discussions of Release 20 and delegates are proposing a new use case, the NWDAF-assisted analytics, detection, and control of User Plane Function (UPF) traffic patterns. The main goal is to mitigate potential abnormal UE or application traffic that may lead to degradation of UPF performance.

The releases also supported evolutions in the NWDAF architecture, starting with a monolithic structure in Release 16. Release 17 introduced a distributed architecture, dividing NWDAF into two components: the Analytics Logical Function (AnLF) and the Model Training Logical Function (MTLF). It also defined new functions to support NWDAF: the Analytics and Data Repository Function (ADRF), the Data Collection Coordination Function (DCCF), and the Messaging Framework Adapt Function (MFAF), as shown in Fig. \ref{fig:NWDAF-arch}. Release 18 brought significant architectural enhancements, including support for roaming, Horizontal Federated Learning (HFL), multi-vendor ML model sharing, and NWDAF performance monitoring (e.g., analytics and ML model accuracy). Finally, in Release 19, SA2 is focusing its efforts on introducing Vertical Federated Learning (VFL) as a feature to NWDAF.

\begin{figure}[ht]
    \centering
    \includegraphics[width=\linewidth]{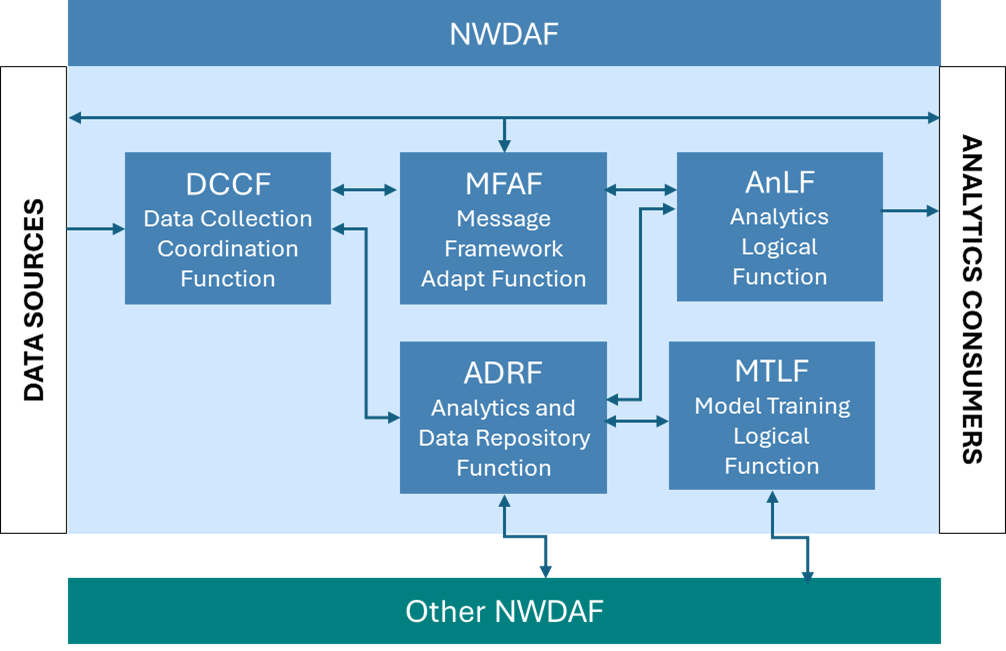}
    \caption{NWDAF Architecture as specified in 3GPP Release 19.}
    \label{fig:NWDAF-arch}
\end{figure}

\section{Related Work \label{sec:related}}

Following 3GPP standardization is not always straightforward. Not only are specifications typically developed approximately five years ahead of actual implementation, but the volume of information and complexity of standards have been increasing substantially every year. Providing the academic community with an overview of ongoing discussions among rapporteurs is a significant and valuable effort that helps align standardization with the state-of-the-art.

Some works in the literature analyze 3GPP standards focused on AI and Data Analytics but fail to simultaneously address advancements in NWDAF use cases, architecture evolution, implementations by researchers, and used datasets. Related works can be cited, as presented in Table \ref{tab:related}. This work offers a contribution to the existing literature by presenting the architecture and use cases through Release 19 and the future directions of Release 20, focusing exclusively on NWDAF. Additionally, it provides comprehensive coverage of existing implementations and datasets employed in related research, and also proposes new use cases.

\begin{table}[hbt]
\caption{Comparison of Related Surveys Regarding NWDAF}
\label{tab:related}
\resizebox{\columnwidth}{!}{%
\begin{tabular}{|c|c|c|c|c|}
\hline
\textbf{Reference} & \textbf{\begin{tabular}[c]{@{}c@{}}3GPP \\ Release\end{tabular}} & \textbf{\begin{tabular}[c]{@{}c@{}}Focus on \\ NWDAF\end{tabular}} & \textbf{Implementations} & \textbf{Datasets} \\ \hline
\cite{duan2021intelligent} & 16 & \ding{119} & \ding{109} & \ding{109} \\ \hline
\cite{koufos2021trends} & 17 & \ding{119} & \ding{119} & \ding{109} \\ \hline
\cite{coronado2022zero} & 17 & \ding{119} & \ding{119} & \ding{119} \\ \hline
\cite{niu2022survey} & 18 & \ding{108} & \ding{109} & \ding{109} \\ \hline
\cite{garcia2023network} & 18 & \ding{108} & \ding{109} & \ding{109} \\ \hline
\cite{sun2024combination} & 19 & \ding{119} & \ding{119} & \ding{109} \\ \hline
This work & 20 & \ding{108} & \ding{108} & \ding{108} \\ \hline
\end{tabular}
}
\newline
\raggedright
\noindent Symbol legend: \ding{108} Full coverage, \ding{119} Partial coverage, \ding{109} Not covered
\end{table}

\section{Methodology \label{sec:methodology}}

A preliminary search for the term ``NWDAF'' within titles, abstracts, and keywords in the Scopus database returned 95 articles, of which 79 were indexed in the IEEE Xplore digital library. To maintain methodological consistency, without losing generality, this study confined its literature retrieval to IEEE Xplore. After a screening process involving both abstracts and full texts, a total of 23 articles were selected for detailed analysis.

These selected studies were systematically examined across key dimensions: implementation approaches, primary use case domains, availability of code and datasets, and the machine learning algorithms employed. This analytical framework facilitated the identification of prevailing trends and highlighted existing research gaps in the current landscape of NWDAF.

\section{Discussion \label{sec:results}}

The analysis presented in Table \ref{tab:discussion} highlights the rapid evolution and experimentation of NWDAF in 5G networks. A clear trend is the increasing use of open-source 5G core projects such as Free5GC\footnote{https://free5gc.org/} for the latest releases of 3GPP and Open5GS\footnote{https://open5gs.org/}, OpenAirInterface (OAI)\footnote{https://openairinterface.org/} for older releases. The Release 16 references in earlier studies reflects the foundational phase of NWDAF exploration, while more recent articles have adopted Release 17 and Release 18, signaling growing maturity and integration of advanced analytics capabilities in the 5G ecosystem. However, no studies have yet registered any use of Release 19 or more.

\begin{table*}[hbt]
\caption{Articles by Release, Testbed, Use Case, Dataset and Code Availability, and ML Algorithm}
\label{tab:discussion}
\centering
\resizebox{.85\linewidth}{!}{%
\begin{tabular}{|c|c|c|c|c|c|c|c|c|c|}
\hline
\textbf{Reference} & \textbf{Release} & \textbf{Testbed} & \textbf{Slice} & \textbf{User} & \textbf{Network} & \textbf{Service} & \textbf{Dataset} & \textbf{Code} & \textbf{ML Algorithm} \\ \hline
\cite{chouman2022towards} & 16 & Open5GS &  &  & \ding{51} &  & \ding{51} &  &  \\ \hline
\cite{manias2022model} & 16 & Open5GS &  & \ding{51} &  &  & \ding{51} &  & LSTM \\ \hline
\cite{manias2022nwdaf} & 16 & Open5GS &  &  & \ding{51} &  & \ding{51} &  & K-Means \\ \hline
\cite{quadrini2023data} & 16 & ANChOR &  & \ding{51} &  &  &  &  &  \\ \hline
\cite{abbas2022ensemble} & 16 & OAI &  & \ding{51} & \ding{51} &  & \ding{51} &  & \begin{tabular}{c}
     GBM, Random Forest, \\
     Catboost, XGBoost 
\end{tabular} \\ \hline
\cite{abbas2021network} & 16 &  & \ding{51} & \ding{51} & \ding{51} &  & \ding{51} &  & \begin{tabular}{c}
     GBM, Random Forest,  \\
     Catboost, LSTM
\end{tabular}  \\ \hline
\cite{sevgican2020intelligent} & 16 &  &  & \ding{51} & \ding{51} &  & \ding{51} &  & \begin{tabular}{c}
     Linear and Logistic Regression,  \\
      XGBoost, LSTM, RNN 
\end{tabular}\\ \hline
\cite{vidhya2020anticipatory} & 16 &  &  & \ding{51} &  & \ding{51} &  &  & Fuzzy \\ \hline
\cite{mekrache2023combining} & 17 & OAI &  & \ding{51} & \ding{51} &  & \ding{51} &  & LSTM, SVM \\ \hline
\cite{nadar2024enhancing} & 17 & OAI &  & \ding{51} &  &  & \ding{51} & \ding{51} & LSTM \\ \hline
\cite{bayleyegn2024real} & 17 & Free5GC &  & \ding{51} & \ding{51} &  &  &  & \begin{tabular}{c}
     Linear Regression, Random Forest,\\
     Decision Tree
\end{tabular}  \\ \hline
\cite{bolla2023open} & 17 & Free5GC &  & \ding{51} &  &  &  & \ding{51} & SARIMA \\ \hline
\cite{de2024nwdaf} & 17 & Free5GC &  & \ding{51} &  &  & \ding{51} & \ding{51} & \begin{tabular}{c}
     Decision Tree, Random Forest,  \\
     MLP
\end{tabular}  \\ \hline
\cite{rajabzadeh2023federated} & 17 & Free5GC &  &  & \ding{51} &  &  & \ding{51} & LSTM \\ \hline
\cite{chen2022meta} & 17 &  &  & \ding{51} &  &  & \ding{51} &  & RNN, LSTM, GRU \\ \hline
\cite{chen2023c} & 17 &  &  &  &  &  &  &  & Decision Tree, SVM \\ \hline
\cite{zhang20245gc} & 17 &  &  &  &  &  & \ding{51} &  & MLP, CNN \\ \hline
\cite{jeon2024hierarchical} & 18 & Free5GC &  & \ding{51} &  &  & \ding{51} & \ding{51} & LSTM \\ \hline
\cite{jeong2024implementation} & 18 & Free5GC &  & \ding{51} & \ding{51} &  &  &  &  \\ \hline
\cite{oliveira2024anomaly} & 18 & Free5GC &  & \ding{51} &  &  & \ding{51} &  & Bagging   Predictor \\ \hline
\cite{kan2024mobile} & 18 &  &  & \ding{51} & \ding{51} &  & \ding{51} & \ding{51} & LLM \\ \hline
\cite{zhang2024fair} & 18 &  &  & \ding{51} &  &  & \ding{51} &  & \begin{tabular}{c}
     AGP, Autoencoder, Random Forest, \\
    OneClassSVM, Isolation Forest
\end{tabular}  \\ \hline
\cite{zhou2023securing} & 18 &  &  &  &  &  & \ding{51} &  & MLP \\ \hline
\end{tabular}%
}
\end{table*}

One of the most prominent observations is the emphasis on user-centric and network-centric analytics. Although some work focuses on predicting user behaviour and improving QoE, others prioritize network-level anomaly detection and performance monitoring. This division reveals the dual role that NWDAF is expected to play: both as a feedback loop for optimizing user services and as a control loop for self-healing and resource optimization in the network.

The application of ML in the surveyed works is notably diverse. Traditional models such as Decision Trees, Random Forests, and XGBoost, often used for anomaly detection, coexist with deep learning techniques, including Long Short-Term Memory (LSTM), Recurrent Neural Networks (RNNs) for time-series prediction, and even transformer-based models like Large Language Models (LLMs) for intent discovering. This heterogeneity reflects the exploratory nature of the field, where the optimal ML approach remains an open research question. These trends underscore the increasing complexity of data sources and highlight the necessity for flexible, adaptive analytics frameworks in NWDAF.

Datasets play a pivotal role in NWDAF research, forming the foundation upon which most studies are built. A combination of real, synthetic, and benchmark datasets is commonly employed, and several works explicitly share their datasets to promote reproducibility.\footnote{https://romerocode.github.io/towards-network-data-analytics-in-5g/} The availability of specific types of datasets often influences the choice of use cases, as researchers tend to prioritize those for which data is more accessible.

For instance, although the slice load level use case has been included in the 3GPP specification since Release 15, only one study to date has attempted its implementation. This underscores how research efforts are often guided by the availability of datasets and the feasibility of simulating specific scenarios. Furthermore, since network slicing is not yet widely deployed in commercial mobile networks, data scarcity remains a major barrier. A study by Ericsson \cite{ericsson} indicates that most enterprises are expected to begin to trial network slicing by 2026, highlighting the gap between standardization and real-world adoption. Consequently, obtaining sufficient and realistic data for experimental purposes continues to be a significant challenge in this domain.

Additionally, the limited sharing of code artifacts reveals a notable gap in the field. None of the most widely used 5G open-source projects have yet officially deployed NWDAF to their official repositories, emphasizing the need for a development of community-driven toolchains to facilitate NWDAF experimentation. Currently, only a small number of projects make their implementations publicly available.\footnotemark[\value{footnote}]

\section{New Proposed Use Cases \label{sec:proposed}}

After an analysis of existing use cases and research gaps, two new use cases are proposed for NWDAF:

\textbf{NWDAF-assisted Charging Bypass Fraud Detection}: Charging bypass occurs when users exploit zero-rated traffic configurations or other vulnerabilities to access services that should be charged \cite{charging-bypass}. NWDAF can help detect such anomalies by continuously analyzing user behaviour patterns, correlating service usage with charging records, and identifying inconsistencies. Using ML models trained on fraud scenarios, NWDAF can identify suspicious activity and trigger alerts or policy adjustments in near real-time.

\textbf{NWDAF Open Gateway Network Interface}: NWDAF act as a centralized analytics engine to mediate interactions between Open Gateway APIs\footnote{https://www.gsma.com/solutions-and-impact/gsma-open-gateway/} and NFs, with a great monetization potential. With the support of MFAF, it can align business-oriented API requests with real-time network data. Telefónica has expressed interest in expanding NWDAF's role to meet Open Gateway use cases\footnote{https://www.telefonica.com/en/mwc/agora-2025/5g-innovation-and-monetization-network-data-analytic-function-nwdaf-potential}, although these initiatives are still exploratory, with no standardized approach or comprehensive studies available. A related idea is explored in \cite{10.1145/3719159.3721225}, where NWDAF supports only performance APIs metrics. The proposed use case goes further by advocating for a standardized integration of CAMARA project\footnote{https://github.com/camaraproject} endpoints with NWDAF, enabling API exposure that is both network-aware and dynamically adaptive.

\section{Conclusion \label{sec:conclusion}}

Through a comprehensive review of 3GPP specifications and the academic state-of-the-art on network data analytics in 5G and beyond, this study identified key trends in NWDAF research and highlighted existing gaps between standardized definitions and real-world implementations. Over 70 papers were initially surveyed, with 23 selected for in-depth analysis regarding implementation, datasets, and ML algorithms. The findings serve as a valuable starting point for new researchers entering the field and propose novel use cases that can be further explored in future works.

\bibliographystyle{ieeetr}
\bibliography{main}

\end{document}